# Mechanism of cellular effect directly induced by magnetic nanoparticles under magnetic fields

Linjie Chen[1,2,3 #], Changyou Chen[1,3 #], Pingping Wang[1,3], Tao Song[1,2,3 *]

1. Beijing Key Laboratory of Bioelectromagnetism, Institute of Electrical Engineering, Chinese Academy of Sciences, Beijing, China;

2. University of Chinese Academy of Sciences, Beijing, China;

3. France-China Bio-Mineralization and Nano-Structures Laboratory, Beijing, China;

[#] **These authors contributed equally to this work.**

**\*Author for correspondence:** Tao Song

**Email:** songtao@mail.iee.ac.cn

**Address for all authors:** No. 6 Bei'er Tiao Zhongguancun HaiDian, Beijing, 100190, China

**Abstract:** The interaction of magnetic nanoparticles (MNPs) with various magnetic fields could directly induce cellular effects. Many scattered investigations have got involved in these cellular effects, analyzed their relative mechanisms and extended their biomedical uses in magnetic hyperthermia and cell regulation. This review reports these cellular effects and their important applications in biomedical area. More importantly, we highlight the underlying mechanisms behind these direct cellular effects in the review from the thermal energy and mechanical force. Recently, some physical analyses showed that the mechanisms of heat and mechanical force in cellular effects are controversial. Although the physical principle plays an important role in these cellular effects, some chemical reactions such as free radical reaction also existed in the interaction of MNPs with magnetic fields, which provides the possible explanation for the current controversy. It's anticipated that the review here could provide readers a deeper understanding of mechanisms of how MNPs contribute to the direct cellular effects and thus their biomedical applications under various magnetic fields.

# 1 Introduction

With the explosive development of nanotechnology, multifarious nanomaterials emerge quickly. As a typical nanomaterials, magnetic nanoparticles (MNPs) contain iron oxides compound, cobalt and nickel compound, or their dopant compound [1-2]. MNPs display their size within dozens to hundreds of nanometers and have a high specific surface area. As magnetic materials, MNPs possess the outstanding properties of magnetism, which can be affected by external magnetic fields. Generally, MNPs are synthesized chemically and further decorated by proteins needed [3]. In addition to this, some biological magnetic nanoparticles (BMPs) also receives considerable attentions, such as magnetosome, a particular kind of $Fe_3O_4$ or $Fe_3S_4$ nanocrystal covered by biological membrane which is formed by magnetotactic bacteria [4-7].

Due to versatile intrinsic properties, MNPs containing magnetosomes have promoted challenging innovations in biomedical application through their interplay with various magnetic fields. In combination with magnetic resonance imaging, MNPs are a fairly effective medical imaging contrast agent [8]. MNPs are also used as drug carrier [9-10] or in targeted therapy, and designed to be biosensors [9, 11]. As an indispensable tool, MNPs play an important role in those applications. During the applications, an external magnetic field is generally required. In addition, MNPs could directly induce the cell effects under magnetic fields. For instance, when exposed in an alternating magnetic field (AMF) MNPs are able to be used in killing tumor cells [12-14]. Recently, the interaction of MNPs and a magnetic field induced mechanical force which is directly applied for the cell regulation [15-16].

With such issues in mind, here in this review, we try to present a brief summary of the cellular effects directly induced by the interaction of MNPs with magnetic fields. We describe the cellular effects of heat induced by MNPs under an alternating magnetic field. Besides the heat effects, this review also addresses the direct effect of mechanical force on cell status utilizing MNPs in a magnetic field. Finally, we discuss the cellular effects of MNPs under magnetic fields from the point of view of free radicals. The related mechanisms behind these cellular effects are simultaneously elaborated.

## 2 Effect of heat

When interplaying with an alternating magnetic field, MNPs transform the field energy into the heat through the mechanism of magnetic hysteresis, Néel or Brown relaxation and eddy current effect [17-19]. Generally, the temperature induced by the technology must increase to 42 ºC at least to harm needless cells [20-21]. The heat effect resulted from MNPs in an alternating magnetic field is also called magnetic hyperthermia which could be used in biomedical therapy.

Cancer is an important treatment target of magnetic hyperthermia. Early magnetic hyperthermia simply injected MNPs into the intratumor and recent reviews have covered the progress [22-23]. Compared to chemotherapy and radiotherapy, the traditional magnetic hyperthermia would not induce the damages of whole body, but still cannot realize targeting to deep lesion location. Targeted magnetic hyperthermia is then developed and considered as a more advanced strategy, where targeted methods depend generally on the receptor-acceptor bind and antigen-antibody reaction. EGF-tagged MNPs were first internalized and when exposed to an AMF they can result in tumor cell death [24-25]. Sanchez C *et al* also used MNPs decorated with Gastrin whose receptors were overexpressed in several malignant cancers to treat Flp-In CCK2R-293 cells and apoptosis and cell death were observed under an AMF [26]. Likewise, MNPs are modified by relevant antibody against specific antigen expressed on the tumor cell surface to facilitate the targeted magnetic hyperthermia, *i.e.* anti-HER2 antibody coated MNPs [27] and anti-Fas antibody-conjugated MNPs [28]. The kind of targeted magnetic hyperthermia could not hurt normal tissue around tumor and allow MNPs to reach in the deep tissue, supporting a novel idea for cancer therapy. Apart from artificially-synthesized MNPs used in magnetic hyperthermia, the kind of biosynthetic magnetosomes inside the cells is also employed to kill cancer cells under an AMF. Magnetosomes could be modified easily due to their biological membrane, and have the feature of good biocompatibility and single magnetic domain. Because of their unique characteristics, magnetosomes produce a larger amount of heat than the smaller superparamagnetic chemically synthesized nanoparticles under the same AMF and hence has a good anti-tumor activity by using magnetic hyperthermia [29-32].

Currently, magnetic hyperthermia has showed its effect against parasitic infections [33]. For pathogenic infection, especially caused by drug-resistant microbe, magnetic hyperthermia is probably an effective way. Kim MH *et al* showed that MNPs conjugated by anti-Protein A antibody were able to target *S. aureus* and could kill the pathogen both *in vitro* and *in vivo* under the AMF [34]. Magnetotactic bacteria which synthesize magnetosomes were also applied to combat *S. aureus*-induced skin infection under an alternating magnetic field [35]. Antibody-coated magnetotactic bacteria could target *S. aureus* because of the affinity between Protein A expressed in *S. aureus* and the Fc fragment of the antibody (Figure 1a). After placed in alternating magnetic field, the length of the wound in mouse tail was reduced significantly, which finally improved the healing effect (Figure 1b and 1c).

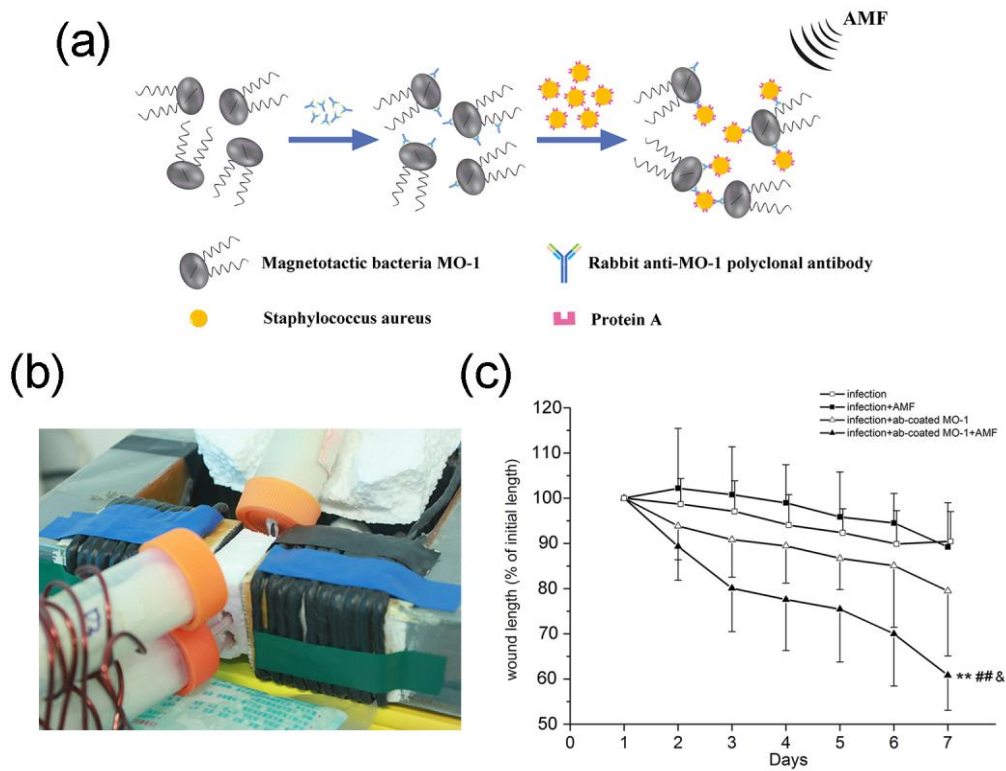

Figure 1 Magnetotactic bacteria-mediated magnetic hyperthermia for killing of *S. aureus*. (a) The schematic representation of magnetotactic-bacterium-mediated magnetic hyperthermia [35]. (b) Application of an alternating magnetic field to *S. aureus*-infected wound in mice tails. (c) The healing effect of the *S. aureus*-infected wound by MO-1-mediated magnetic hyperthermia [35].

For both traditional and targeted magnetic hyperthermia, abundant MNPs are used and then cause temperature rise in the bulk solution under an AMF. The temperature rise probably is the major reason for tumor or bacterial cell death. Particularly, targeted MNPs are probably phagocytized by tumor cells, coated by vesicles and transferred to the lysosomes. Temperature rise by the MNPs under an AMF could result in the change of lysosomal membrane permeabilization or immediate destruction of lysosomes [24, 26]. Under the condition, such factors from lysosomes as chymotrypsin and cathepsin were released to cytoplasm and then activated cell apoptosis [36-37].

When slight MNPs are used under an AMF to induce heat, a perceptible macroscopic temperature rise is not monitored. However, the direct cellular effect in a context of this sort is still confirmed to be effective [25, 33, 38-39]. So what does the cellular effect attribute to in the condition? Based on the development of biotechnology and nanotechnology, the temperature around MNPs was investigated and measured. Through different methods, those studies showed that the temperature on the surface of MNPs is higher than that in the solution, although the local high temperature decays exponentially with increasing distance [40-44]. From the point of view, the local high temperature around MNPs could damage cell or organelle membrane and cause cell apoptosis or death, and thus may be a major factor for cellular effects especially in targeted hyperthermia.

Considering the local temperature rise around MNPs under an alternating magnetic field, on the other hand, some scholars tried to use local high temperature to induce cellular effects, such as regulation of cell behaviors. They first attached MNPs to a temperature-sensitive ion channel and then applied an AMF to generate a local temperature rise around MNPs. Subsequently, the temperature-sensitive ion channel could be activated by the high temperature. When transient receptor potential A1 (TRA1), a temperature-sensitive cation channel, was conjugated by MNPs and then placed in a radio-frequency magnetic-field, the channel was activated by heat [40]. Chen R *et al* targeted MNPs to a transient receptor potential vanilloid-1 (TRPV1), and found that the nanoparticles triggered widespread and reversible firing of TRPV1$^+$ neurons when interacting with an AMF [45]. Further, the interplay of iron oxide nanoparticles with an AMF was applied for deep brain stimulation[45] and regulation of plasma glucose in mice [46]. The paramagnetic protein ferritin was recently used to substitute for MNPs. Stanley SA *et al* developed a genetically encoded system where GFP-tagged ferritin associated intracellularly with a camelid anti-GFP-TRPV1 fusion protein, which can activate channel by the interaction of ferritin with noninvasive low-frequency radio waves [47]. They also demonstrated that the system could be used to lower blood glucose by remotely stimulating insulin transgene expression with a radio-frequency magnetic-field [47]. Later the same group improved the genetically encoded system for non-invasive, temporal activation or inhibition of neuronal activity *in vivo* by radio frequency treatment and described its use to study central nervous system control of glucose homeostasis and feeding in mice under [48]. These studies offer us the new tools to regulate cell functions.

However, the controversy over the temperature gradient around MNPs under an AMF existed as well. Some theoretical and experimental analyses showed that the temperature rise on the surface of MNPs was negligible [49-52]. Moreover, Meister M calculated the thermal energy induced by Ferritin under radio-frequency magnetic field and found that either temperature increase or gradient cannot activate TRPV1 [53]. There might also be other mechanisms in the process.

## 3 Effect of mechanical force

Due to magnetic properties, MNPs are able to create mechanical force or torque when interacting with magnetic fields. If MNPs conjugate to the cell membrane or are endocytosed into cells, mechanical force could directly induce cellular effect.

Generally, magnetic hyperthermia employs MNPs with the property of superparamagnetism and isotropy. Upon applying an inhomogeneous magnetic field, MNPs will display the phenomenon of oscillation. The behaviors can generate shear forces which may make cells broken directly or cause the incompleteness of organelles that induced cell apoptosis. Carrey J *et al* theoretically studied and showed MNPs oscillate mechanically in an inhomogeneous AMF with the great possibility of creating shear force [54]. The theoretical results also showed that oscillating MNPs may generate ultrasound [56]. Ultrasound is proved to give rise to cell apoptosis or dissolution [86-87]. Besides, superparamagnetic and isotropous MNPs are also placed in other magnetic fields to create mechanical force which directly induces cellular effects. Zhang E *et al* demonstrated that shear forces was generated by MNPs in an unique dynamic magnetic field, which further induced cell death [55]. With regard to the effect induced by rotating MNPs under the magnetic field, Yue TT *et al* theoretically investigated the interaction mechanism between lipid membranes and rotating MNPs through simulation and found that the rotating nanoparticles promoted cell uptake and also rupture the mechanical membrane [56].

Anisotropic MNPs are extensively exploited to induce cellular effect by generating mechanical force under various magnetic fields as well. Kim DH *et al* first used microdiscs that possess a spin-vortex ground state to interface with cells; when an alternating magnetic field is applied, the microdisc vortices shift, creating an oscillation which transmits a mechanical force to destruct cancer cell [57]. They showed that the spin-vortex-mediated stimulus creates two dramatic effects: compromised integrity of the cellular membrane, and initiation of programmed cell death. Cheng Y *et al* destructed glioma cells by applying spin-vortex, disk-shaped permalloy magnetic particles in a low-frequency, rotating magnetic field which could create a strong mechanical force and induce programmed cell death [58]. Cheng D *et al* used rod-shaped MNPs to induce HeLa cell viability decrease by about 2-fold than spherical MNPs after being exposed to AMF [59]. A swing magnetic field with low frequency and low heat-production were designed in order to evaluate the torque effect of magnetotactic bacteria on *S. aureus* (Figure 2a and b) [60]. Under the swing magnetic field, the addition of magnetotactic bacteria in suspension of *S. aureus* caused negligible temperature increase (Figure 2c), and *S. aureus* was killed only when attached to magnetotactic bacteria (Figure 2d) [60]. Finally, we analyzed and calculated the mechanical force created by magnetotactic bacteria under the swing magnetic field with a value of 8.3kPa [60]. According to previous reports [61-62], the mechanical force may be enough to induce the damage of *S. aureus* or directly kill the pathogen.

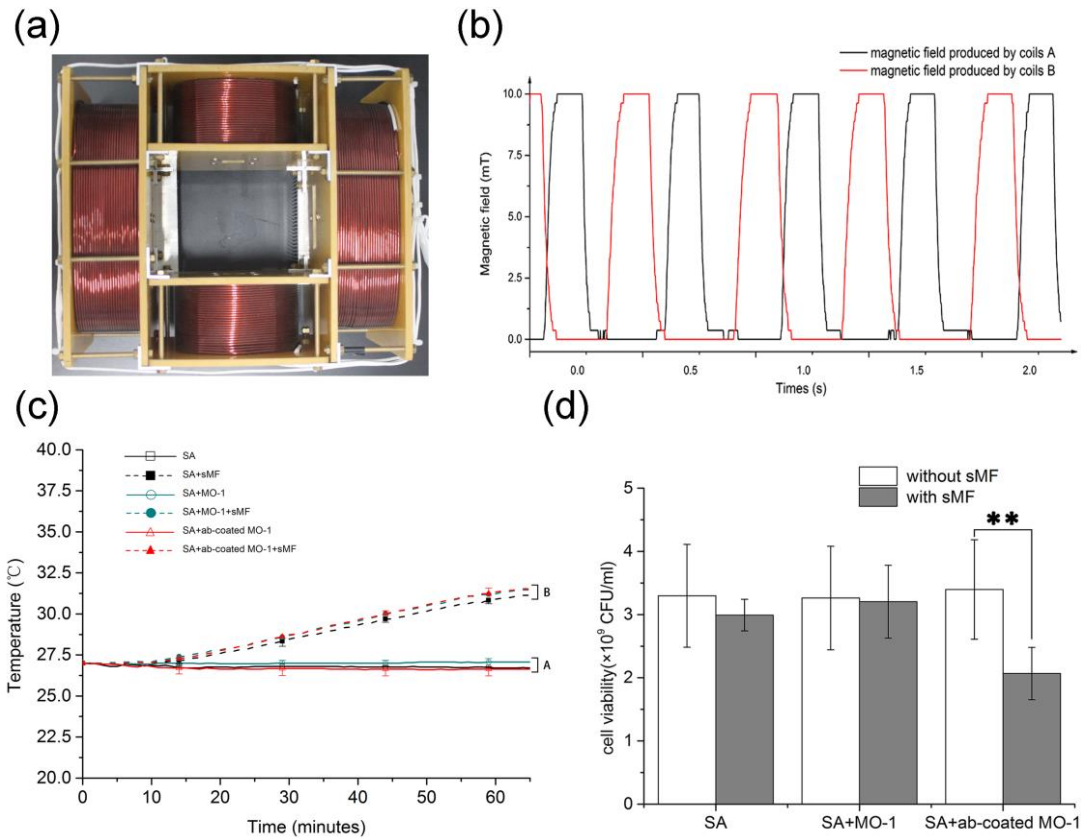

Figure 2 Killing of *S. aureus* by magnetotactic bacteria in a swing magnetic field. (a) The coils of the swing magnetic field generator. (b) The swing magnetic field with the intensity of 10 mT and the frequency of 2 Hz. (c) The temperature measured in *S. aureus* alone suspensions, the mixed suspension of *S. aureus* with free MO-1 cells, the attached suspension of antibody-coated MO-1 cells to *S. aureus* with or without the swing magnetic field. (d) The killing effect of magnetotactic bacteria under the swing magnetic field on *S. aureus*. Panels b, c and d are reproduced with permission from [66]. Copyright © 2017 Copyright Clearance Center, Inc.

Actually, the mechanical force from the interaction of MNPs and magnetic fields has been used for controlling cell status. By applying a gradient magnetic field, the magnetic nanoparticles get a drag force (Figure 3a). When the nanoparticles bind to ion channels, the drag force may switch on the ion channel. Hughes S *et al* showed that manipulation of MNPs that targeted against a mechanosensitive ion channel TREK-1 by a gradient magnetic field, could lead to the change of TREK-1 activity and the whole-cell currents [63]. Lee JH *et al* proved that the cube-shaped MNPs bound to components of cellular membranes and that their interplay with a magnetic field from an electromagnet can exert mechanical force on the cells, inducing the influx of ions into the hair cell [64]. Magnetic force stimulation with ferromagnetic nanoparticles under a gradient magnetic field was also found to trigger calcium influx in cortical neural networks [65]. Different from the drag force, the torque of MNPs from the change of direction of the applied magnetic field was also used to control cells (Figure 3b). Here an important thing to note is that when the superparamagnetic nanoparticles are used, a magnetized field should be first applied to magnetize the kind of MNPs. After applying a strong external magnetic field to magnetize the modified microbeads, Wang N *et al* used the magnetized microbeads under a 90º-oriented magnetic field to create a torque that could the deformation of cell cytoskeleton [66]. Chowdhury F *et al* also used a strong magnetic impulse to magnetize the microbeads that attached to mouse embryonic stem cells and found that microbeads promoted early development of stem cells via applying a twisting field [67]. Besides, aggregation of MNPs may induce the cluster of MNPs-attached receptors under a magnetic field and then initiate the signal transduction, which is another approach to regulate cells (Figure 3c). Cho MH *et al* applied a static magnetic field to aggregate MNPs bound death receptor 4, which then promotes apoptosis signaling pathways [68].

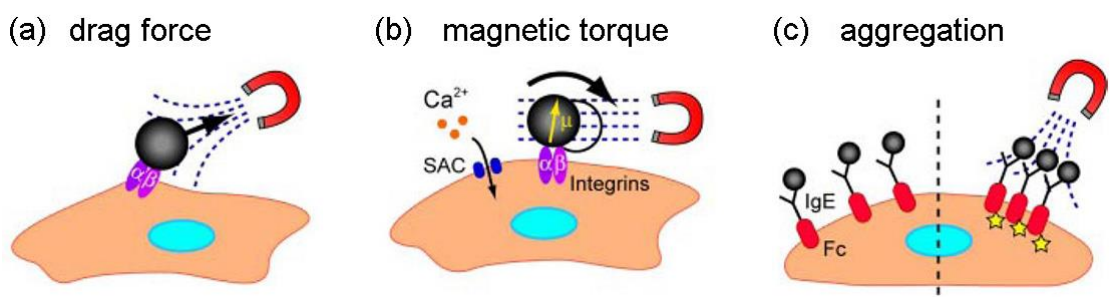

Figure 3 Mechanical force from the interaction of MNPs and magnetic fields for controlling cell status. (A) By applying a gradient magnetic field, the magnetic nanoparticles get a drag force used for the control of cells. (B) The mechanical torque induced by MNPs from the change of direction of the applied magnetic field was also used to control cells. (C) Aggregation of MNPs may induce the cluster of MNPs-attached receptors under a magnetic field and then initiate the signal transduction. Copyright © 2010, Endocrine Society.

Recently, the iron-sulfur cluster assembly protein 1 or ferritin have been used to regulate cells by interplaying with magnetic fields. Different from the attachment of MNPs to ion channels via chemical methods, iron-sulfur cluster assembly protein 1 or ferritin is co-expressed with ion channels through genetic tools to form a fusion proteins. Then an external magnetic field is applied to induce cellular effect. Long XY *et al* expressed an iron-sulfur cluster assembly protein 1 in HEK-293 cells and cultured hippocampal neurons; and showed application of the external magnetic field resulted in membrane depolarization and calcium influx [69]. Wheeler MA *et al* fused the cation channel TRPV4 to ferritin to form a single-component, magnetically sensitive actuator, "Magneto," and validated noninvasive magnetic control over neuronal activity using the "Magneto" [70].

MNPs with different properties can generate mechanical force or torque under different kinds of magnetic fields. On the one hand, the mechanical force or torque could induce cellular effects by directly destructing cell membrane or lysosomal membrane. On the other hand, the mechanical force or torque changes the condition of the cytoskeleton, intracellular ion channel, mitochondria or other signal transduction pathway. Followed is that cells may experience different status, such as growth, differentiation or death. Particularly, the abnormal conditions of ion channel trigged by mechanical force，*i.e.* the abnormal switch on-off, excessive open or close can induce cell apoptosis and death (see in review [71-73]). It provides us with a new thought for cell regulation or diseases therapy. However, the theoretical calculation from Meister M [53] showed that mechanical force from the interaction of ferritins with different kinds of magnetic field was unable to gate the ion channels. Insufficient consideration of factors in live cells may be a reason that lead to the contradiction. Moreover, besides the heat and mechanical force, the interaction of MNPs with magnetic fields may result in chemical reactions such as free radical reaction. The kind of the undetermined mechanism is also an important reason.

## 4 Effect of free radical

Free radicals such as reactive oxygen species (ROS) are the product of cellular metabolism and participate in physiological functions. However, the over-generation of free radicals seriously impact normal cell activity. Domenech M *et al* demonstrated that lysosomal membrane permeabilization induced by MNPs under an AMF is correlated with the production of ROS [24]. Further, Connord V *et al* recorded AMF-induced production of ROS in cells having incorporated DY647-MNPs-gastrin in a real-time form and found that ROS level within cells submitted to the AMF increased significantly compared to the cells outside the AMF [74]. Upon application of a static magnetic field, MNPs exert synergistic adverse effects such as reduced cell viability, apoptosis, and cell cycle aberrations on hepatocytes *in vitro* and *in vivo*; and the apoptotic effect was dependent on levels of ROS [75]. Combined exposure of extremely low frequency magnetic field (50 Hz, 400 μT) and MNPs-SiO2 resulted in remarkable cytotoxicity and increased apoptosis in the differentiated rat pheochromocytoma cell [76]. Superabundant free radicals attack almost all of macromolecules including DNA, protein, carbohydrate and lipids and hence are harmful to cells. On the contrary, MNPs may decrease intracellular free radicals under magnetic fields. A study showed that MNPs could significantly decrease in $H_2O_2$-mediated oxidative stress in combination with an external electromagnetic field and act as free radical scavenger in the cure of spinal cord injury [77]. Shin J *et al* also demonstrated that BMPs enhanced cell growth and have anti-apoptotic effect in a static magnetic field [78]. However, they attributed the positive effect to the factor that the interplay of BMPs with external static magnetic field within the cell may induce the alteration of the diamagnetic anisotropy of membrane phospholipids, resulting in modulation of cell signaling related to the membrane protein and the enhanced cell growth.

Thus, based on currently available data, it seems likely that MNPs could influence the different cellular effect by free radicals. On the one hand, magnetic nanoparticles could play a role in catalytic ROS production. Through the release of $Fe^{2+}$, MNPs take part in the production of ROS by Fenton reaction. It's dramatically noted that when exposed under a magnetic field, MNPs may promote the process. Wydra RJ described the accelerated generation of free radicals by MNPs in the presence of an AMF [79]. Binhi V also theoretically analyzed the generation of free radicals by MNPs in a static magnetic field and found that MNPs can increase the rate of free radical formation [80]. On this occasion, magnetic field generated by MNPs around themselves are orders of magnitude greater than the applied magnetic field [76, 81]. The high magnetic field around MNPs may impact radical pair spin states, resulting in the decrease of recombination rates of radical pair particles [82]. As a result, the free radicals increase. Of course, some argued that the heat stress created by MNPs under an AMF is likely to induce the production of free radicals [83-84]. On the other hand, $Fe_3O_4$ nanoparticles were actually found to have the catalytic activity of peroxidase and catalase, which was called nanozyme [85-86]. The catalytic activity is supposed to be a scavenger that eliminates ROS in the cell. The nanozyme alone promotes cell growth, which was attributed to the MNPs' peroxidase-like activity [87]. Overall, the direct cellular effect of free radicals possibly offers new clues for disease treatment, especially killing of tumor cells. Yet the cellular effect of free radicals induced by MNPs under magnetic fields and their underlying mechanism should be studied deeply. At the same time, whether MNPs as nanozyme are influenced by magnetic fields is not clear and also needs further investigations.

## 4 Conclusion and future perspectives

In this account, we reviewed concisely the direct cellular effect of magnetic nanoparticles under different kinds of magnetic fields. We introduced the first cellular effect of magnetic nanoparticles under an alternating magnetic field which could be used in tumor therapy and infection treatment. The mechanism of cellular effect depends on the heat or temperature gradient around MNPs. Then another direct cellular effect was discussed particularly in cell regulation because the interaction of MNPs with a magnetic field could create the mechanical force. The mechanical force could influence the ion channels or damage the cell membrane and further induce cell apoptosis or death. Meanwhile, it's found that mechanical force might involve in the cellular effect caused by heat. Finally, the effect of free radicals generated by MNPs when interplaying with a magnetic field was analyzed. Although some current understanding of cellular effects directly induced by MNPs in magnetic fields was shown here, some questions about how MNPs could directly induce cellular effect are still indistinct. The mechanisms of the cellular effect of heat and the relationship of MNPs' peroxidase-like activity with magnetic fields need to be confirmed. Therefore, more studies and experimental exploration into these mechanisms are quietly required and can further lead to significant advances in various area of in-depth biomedical applications of MNPs under magnetic fields such as precision medicine of diseases, cell or nerve regulation.

## Acknowledgment

This work was supported by the National Natural Science Foundation of China (NSFC) (31300689).